\documentclass[journal,10pt,twocolumn]{IEEEtran}
\usepackage{graphicx, times, amsmath, amsfonts, comment,amsthm}
\usepackage{amssymb, enumitem, epstopdf, soul, color, cite}
\usepackage[noend]{algorithmic}
\usepackage{algorithm, array}
\usepackage{epstopdf}
\usepackage{subfig}
\usepackage{graphicx}

\newcommand{\beq}{\begin{equation}}
\newcommand{\eeq}{\end{equation}}

\newcommand{\tit}{\textit}

\theoremstyle{plain}

\theoremstyle{plain}

\theoremstyle{plain}

\theoremstyle{plain}

\theoremstyle{plain}

\theoremstyle{plain}

\newcommand {\Ebb}{\mathbb{E}}

\newcommand {\Vbb}{\mathbb{V}}

\begin{document}
\title{Level-Triggered Harvest-then-Consume Protocol with Two Bits or Less Energy State Information}
\author{\IEEEauthorblockN{Sudarshan Guruacharya, Vandana Mittal, and Ekram Hossain \\ \thanks{The authors are with the Department of Electrical and Computer Engineering at the University of Manitoba, Canada (Emails: \{Sudarshan.Guruacharya, Ekram.Hossain\}@umanitoba.ca, mittalv@myumanitoba.ca). The work was supported by a CRD grant (CRDPJ 461412-13) from the Natural Sciences and Engineering Research Council of Canada (NSERC).}} 
}
\maketitle

\begin{abstract}
We propose a variation of harvest-then-consume protocol with low complexity where the harvest and consume phases change when the battery energy level reaches certain thresholds. The proposed protocol allows us to control the possible energy outage during consumption phase. Assuming that the battery is perfect and that the energy arrival is a renewal process, we analyze the duty cycle and the operating cycle speed of the protocol. The proposed protocol also allows for limited battery energy state information. The cases when the system has two-bits, one-bit, and zero-bit  of battery energy state information are studied in detail.  Numerical simulations verify the obtained formulas.
\end{abstract}


\section{Introduction}
\label{sec:intro}

Harvest-then-consume protocol is an instance of time-switching architecture of energy harvesting communication, where harvest and consume phases alternate between each other. In much of the work on this protocol, the duration of a harvest-consume cycle is assumed to be fixed, and the duty cycle is assumed as the control variable, with the goal of maximizing the system throughput \cite[references therein]{Ku2016}. However, this can result in an energy outage when the harvested energy is insufficient to power the consumer's application.

In this letter, we consider a less complex variation of harvest-then-consume protocol  where the phase change happens when the battery energy level crosses certain threshold; hence the qualifier \tit{level-triggered}. The protocol we describe has a built-in guarantee over energy outage. A similar protocol was used for empirical work in \cite{Lee2011} and for relay system in \cite{Krikidis2012,Nasir2015,Bi2016,Gu2016}. All these papers had slotted time process, making the energy inter-arrival time deterministic. In \cite{Krikidis2012,Bi2016,Gu2016}, the energy evolution is modeled as discrete time, discrete state Markov chain and the stationary distribution of the Markov chain is obtained. These papers did not investigate the duty cycle nor the cycle speed of the protocol. In \cite{Nasir2015} duty cycle was studied; but the authors missed the fact that the duty cycle converges to a limiting value for large thresholds. This paper generalizes these works since, in our work, the energy inter-arrival time and energy packet size can be any distribution with finite mean and variance.

Being level-triggered also means that at most two bits of energy state information (ESI) are needed to monitor the changes in the battery (or super-capacitor) level. This is an important observation, missed by prior papers, since (a) the battery's state-of-charge cannot be directly measured, and (b) the accurate estimation of the state-of-charge remains a difficult task, primarily because of limited battery models and parametric uncertainties in these models \cite{Li2017,Zhang2017}. The proposed protocol greatly simplifies the system design, since the dedicated circuitry and complex algorithm needed to monitor the battery's energy state can be reduced or done away with completely, in favor of inexact data. Also, since the goal of the protocol is to ensure energy sufficiency for a given time frame, complicated optimization problem over finite or infinite time horizon is avoided.  Lowering the ESI (two bits or less) can imply cheaper design, lower circuit power consumption, greater robustness to circuit failures, and lower circuit complexity. All these features are important when we design wireless sensor nodes. However, this can  come at the cost of reduced efficiency. Counter-intuitively, we will show that when the threshold energy is large, performance for all the limited ESI cases converge to same value.

Our contributions in this paper are to  (1) analyze the operating speed and the duty cycle of the proposed protocol, and establish upper bounds on performance, (2) for cases when two-bits, one-bit, and zero-bit of ESI is available, given the energy and information outage constraints, and (3) show that they converge to the same limiting value for large threshold. For such analysis, the information of  the recharge time distribution is required \cite{Guruacharya2017}. While these metrics are independent of the purpose of the energy consumption, nevertheless, we will assume that the energy is used to transmit messages in a point-to-point wireless communications system. The case of one-bit ESI and zero-bit ESI are interesting in itself, since they represents  ``half blind'' and ``full blind'' design, respectively. For one-bit ESI, the harvest duration is deterministic; while for zero-bit ESI, both the harvest and consume durations are deterministic. Note that such a simple harvest-then-consume protocol will be particularly suitable for low-complexity energy-harvesting wireless sensor nodes and internet-of-things (IoT).  \tit{To the best of our knowledge, we are not aware of any prior work that deals with limited ESI.}

\section{System Model}
\label{sec:sys-model}

Let $U(t)$ be the energy level of the battery at any given time $t$.  At any given time,  the battery can be in one of the three possible states: (i) Battery is empty, $U(t) = 0$, (ii) Battery is not empty but the energy level is below some required threshold $u$, such that $0 < U(t) < u$, (iii) Battery energy level is above the required threshold, $U(t) > u$.  These three possible energy states can be represented by just \tit{two bits} of information. Thus, we have a system with limited ESI. 

In this letter, we will consider the following simple version of the \tit{harvest-then-consume protocol}:
\begin{itemize}
\item The system switches off and goes into harvest phase when the battery is empty, i.e. after $U(t) = 0$. 
 \item The system switches on and goes into consumption phase after the battery has acquired $u$ amount of energy, i.e. after $U(t) > u$.
\end{itemize}
Here, harvest and consume phases are triggered when battery attains certain fixed levels, hence the name \tit{level triggered}.

For simplicity, during the consumption phase we will assume that the rate of energy consumption (or the consumed power) $p$ is constant. Let $\tau_c = \inf\{t: U(t) > u, U(0) = 0\}$ be the recharge duration and $\tau_d = \inf\{t: U(\tau_c + t) = 0\}$ be the discharge duration. Then, the total harvest-consume cycle duration is $T = \tau_c + \tau_d$. The performance metric of the harvest-then-consume protocol is taken to be 
\beq
\omega =\frac{1}{\Ebb[T]}  \quad \mathrm{and} \quad \rho = \frac{\Ebb[\tau_d]}{\Ebb[T]}. 
\label{eqn:performance-metrics}
\eeq

Here, $\omega$ represents the cycle speed at which the protocol can operate and  has the units of Hertz. The other metric, $\rho$, is the duty cycle, such that $\rho \in (0,1)$. It represents the fraction of time that the system does some useful work and is a dimensionless number. In this paper, we will derive formulas for these two metrics of interest. 

During the harvest phase, we model that the recharge process for perfect battery (i.e. linear charging and no self-discharge) as 
 \beq 
U(t) = \sum_{i=1}^{N_A(t)} X_i,
\label{eqn:recharge-process}
\eeq 
where $N_A(t) = \min\{k: A_0 + A_1 + \cdots + A_k \leq t\}$ counts the number of energy arrivals, $A_{i\geq1}$ is the inter-arrival time of energy packets, $A_0$ is the residual time, and $X_i$ is the energy packet size.  The energy arrival is assumed to be a delayed renewal process. The $\{A_{i\geq 1}\}$ and $\{X_i\}$ are assumed to be independent and identically distributed with finite mean and variance. Also, we assume that $\{A_i\}$ and $\{X_i\}$ are independent of each other. Lastly, we assume that the joint distribution of the random vectors $\{(A_i, X_i) \}$ are identically distributed as $(A,X)$. For notational convenience, we will denote $\lambda = 1/\Ebb[A]$ and $\bar{X} = \Ebb[X]$.

With less than two bits of ESI, the system may not be able tell if the battery has the desired energy level. As such, we need to impose a statistical guarantee on the energy outage. Let the energy outage constraint at the switching  time $t_c$ be 
\beq 
P(U(t_c) \leq u) = \theta_1, 
\label{eqn:energy-outage}
\eeq
where $\theta_1 \in (0,1)$. Here $t_c$ is a fixed duration of recharge, after which the system is turned on for the consume phase. 

Given the recharge process in (\ref{eqn:recharge-process}), 
the mean and variance of $\tau_c$ for large $u$ are, respectively\footnote{Here, $f(x) \sim g(x)$ if and only if $\lim_{x \to \infty}\frac{f(x)}{g(x)} = 1$.} \cite[Eqns. (8), (9)]{Guruacharya2017},
\beq
\Ebb[\tau_c] \sim C_1 + \frac{u}{\lambda \bar{X}}, \quad \mathrm{and} \quad \Vbb[\tau_c] \sim C_2 + \frac{\gamma^2 u}{\bar{X}^3}.
\label{eqn:mean-variance-recharge-time}
\eeq

Here the constants 
$C_1 = \frac{\lambda \gamma^2}{2 \bar{X}^2}$ and $C_2 = \frac{\mu_A^{(3)}}{3\mu_A} - \left(\frac{\mu_A^2 + \sigma^2_A}{2 \mu_A}\right)^2$, where $\mu_A^{(3)}$ is the third moment of $A$; $\sigma_A^2$ and $\sigma_X^2$ are the variances of $A$ and $X$, respectively; and $\gamma^2 = \lambda^{-2} \sigma_X^2 + \sigma_A^2 \bar{X}^2$. For large $u$, we can neglect the constant term and simply write $\Ebb[\tau_c] \sim u/\lambda \bar{X}$ and $\Vbb[\tau_c] \sim \gamma^2 u/\bar{X}^3$. Thus, invoking the central limit theorem, for large $u$ the distribution of $\tau_c$ is given by \cite[Eqn. (11)]{Guruacharya2017}, 
\beq 
P(\tau_c(u) \leq t_c) \approx \Phi\left(  \frac{t_c - \Ebb[\tau_c] }{\sqrt{\Vbb[\tau_c]}} \right),
\label{eqn:recharge-time-distr}
\eeq
where $\Phi(\cdot)$ is the standard normal distribution. 

If the harvested energy is used to transmit information\footnote{Depending on the number of ESI bits, cycle time can be random; thus,  asynchronous communication may be needed.} in a narrow band channel, then in the transmit phase, assuming a point-to-point wireless communications system with transmit power $p$, flat fading channel gain $g$, and additive white Gaussian noise with power $N$, we have the signal-to-noise ratio (SNR) given by $Z = gp/N$. We assume that the transmitter always has data to transmit. 
Let the SNR outage constraint be
 \beq 
 P(Z \leq \zeta) = \theta_2, 
 \label{eqn:info-outage}
 \eeq 
 where $\theta_2 \in (0,1)$ while $\zeta$ is the threshold SNR required for correct decoding of the message signal. Substituting the expression for SNR in (\ref{eqn:info-outage}),  we have $P\left(g \leq \frac{\zeta N}{p}\right) = \theta_2$. Since $P\left(g \leq \frac{\zeta N}{p}\right) = F_G \left(\frac{\zeta N}{p}\right)$, where $F_G$ is the distribution of $g$, we can solve for $p = \frac{\zeta N}{F^{-1}_G(\theta_2)}$.

The metrics $\omega$ and $\rho$ are clearly related to the throughput of the system. If a single symbol is transmitted in every time frame $T$, then the average symbol rate of the system is $1/\Ebb[T] = \omega$. Also, let the symbol energy be $u = p T_s$, where $T_s$ is the fixed symbol duration. Then, for large $u$, we have $\omega \sim \frac{p\rho}{u}$ and $\rho \sim \frac{\lambda \bar{X}}{\lambda \bar{X} + p}$ (as we will see in later sections); and the symbol rate of the system becomes $\omega \sim \frac{\rho}{T_s}$, which is affected the duty cycle $\rho$. This also explains why the duty cycle can be taken as the efficiency of the system. If $\rho = 0$, then no symbol can be transmitted. If $\rho = 1$, then the system transmits at its maximum possible rate. Thus the system is limited by the harvest delay, rather than channel capacity.


\section{With Two Bits of Energy State Information}
\label{sec:two-bit}

With two bits of ESI, the system can know when the battery is empty and when it has sufficient energy. The duration that the recharge process takes to cross the desired energy level $u$ is $\tau_c$, where $\tau_c$ is a random variable. Once the required energy level has been crossed, the system is turned on. The time it takes to fully discharge the battery is $\tau_d = U(\tau_c)/p$; and the total charging and discharging time is $T = \tau_c + \frac{U(\tau_c)}{p}$. Here again $T$ is a random variable. Also, at the level crossing time $\tau_c$, $U(\tau_c) = u + V$, where $V \geq 0$ is the value by which $U(\tau_c)$ overshoots the required energy level $u$. Since $U(t)$ is renewal process, the overshoot is given by the stationary residual density of $X$, assuming $u$ is large, as $f_V(v) = \bar{X}^{-1}[1 - F_X(v)]$. Thus, we have $T = \tau_c + (u+V)/p$. Here, $\tau_c$ and $V$ are independent of each other, thus the distribution of their sum can be obtained by the convolution of their distributions. We can find the mean value of $U$ as $\Ebb[U(\tau_c)] = u + C_3$, 
where $C_3 = (\sigma_X^2 + \bar{X}^2)/2\bar{X}$ is the mean of $V$ which does not depend on $u$. Also, using (\ref{eqn:mean-variance-recharge-time}) 
and the mean of $U(\tau_c)$, we have the mean of $T$ as $\Ebb[T] = \left( \frac{1}{\lambda \bar{X}} + \frac{1}{p} \right)u + \left(C_1 + \frac{C_3}{p}\right)$.

From these, the duty cycle $\rho = \Ebb[\tau_d]/\Ebb[T] = \Ebb[U(\tau_c)]/p\Ebb[T]$ and the system speed $\omega = 1/\Ebb[T]$ are 
\begin{align} 
\rho &= \frac{u+C_3}{ \left(1 + \frac{p}{\lambda \bar{X}}\right)u + ( p C_1 + C_3 )}, \label{eqn:two-bit-rho} \\
\omega &=  \left[\left(\frac{1}{p} + \frac{1}{\lambda \bar{X}}\right)u + \left( C_1 + \frac{C_3}{p}\right) \right]^{-1}. \label{eqn:two-bit-omega} 
\end{align}

As a special case, as $u \to 0$, we have the duty cycle as $\rho = C_3/(pC_1 + C_3)$, and the system's cycle speed as $\omega = p/(pC_1 + C_3)$. This is also the fastest speed that the system can attain; thus $\omega \leq p/(pC_1 + C_3)$. 

Likewise, as $u \to \infty$, we can ignore the constant terms. Thus, $\Ebb[U(\tau_c)] \sim u$ and $\Ebb[T] \sim \left(\frac{1}{p} + \frac{1}{\lambda \bar{X}}\right)u$. In other words, larger the required energy, more we need to wait. Also, the duty cycle is $\rho \sim \frac{\lambda \bar{X}}{\lambda \bar{X} + p}$, and the system's cycle speed is $\omega \sim  \frac{p\rho}{u}$.

Interestingly, this limiting value of $\rho$ is not equal to its value at $u=0$. Setting $u=0$ represents an \tit{opportunistic} scheme where the harvested energy is immediately consumed.  When $u=0$, we have  $U(\tau_c) = X$ and $\tau_d = U(\tau_c)/p = X/p$. Hence, $\Ebb[\tau_d] = \bar{X}/p$. Similarly, $\Ebb[T] = \Ebb[\tau_c] + \Ebb[\tau_d] = C_1 + \bar{X}/p$. Therefore, $\rho = (1 + pC_1/\bar{X})^{-1}$ and $\omega = (C_1 + \bar{X}/p)^{-1}$. For small values of $u$, these formulas for $\rho$ and $\omega$ will not be accurate, since we assume stationary residual distribution for $A_0$ and $V$, which is valid only for large $u$. Note that when $u=0$, only one bit is required to check the battery status; thus this analysis is valid for Section \ref{sec:one-bit} as well.


\section{With One Bit Energy State Information}
\label{sec:one-bit}

Here, we assume that the system can discern whether or not the battery is empty. As such,  during the charging process, only statistical guarantee (\ref{eqn:energy-outage}) can be given for the energy outage. Using (\ref{eqn:mean-variance-recharge-time}) 
and (\ref{eqn:recharge-time-distr}) in (\ref{eqn:energy-outage}), the switching time $t_c$ is given by
\beq 
t_c = C_1 + \Phi^{-1}(1 - \theta_1) \sqrt{C_2 + \frac{\gamma^2 u}{\bar{X}^3}} + \frac{u}{\lambda \bar{X}}.
\label{eqn:switch-on-time}
\eeq
The minima at $u=0$ is $t_{c,\min} = C_1 + \Phi^{-1}(1 - \theta_1) \sqrt{C_2}$, which gives the minimum waiting time for an energy packet to arrive.

Once the system is switched on at $t_c$, the duration it takes for the battery to completely discharge is $\tau_d = U(t_c)/p$. We can find the distribution for the discharge duration as \cite{Beichelt2002}
\beq 
P(\tau_d \leq t_d) = P(U(t_c) \leq p t_d) = \Phi \left( \frac{pt_d - \lambda \bar{X} t_c}{ \gamma \lambda^{3/2} \sqrt{t_c}} \right). 
\label{eqn:discharge-time-distr}
\eeq
Thus, the mean discharge time is $\Ebb[\tau_d] = \lambda \bar{X} t_c/p$.

Since the system can detect when the battery is empty, we can start the recharging process when the battery is completely discharged. Thus, the cycle duration is $T = t_c + \tau_d$. The distribution of $T$ is  
\[ P(T \leq t) = P(\tau_d \leq t - t_c) = \Phi \left( \frac{pt - (p + \lambda \bar{X}) t_c}{ \gamma \lambda^{3/2} \sqrt{t_c}} \right). \]
Thus, the mean of $T$ is $\Ebb[T] =  (p + \lambda \bar{X}) t_c/p$. Hence, from these the duty cycle and the cycle speed are 
\begin{equation}
\rho  = \frac{\lambda \bar{X}}{p + \lambda \bar{X}}, \quad \mathrm{and} \quad \omega = \frac{p}{(p + \lambda \bar{X}) t_c}.
\label{eqn:one-bit-rho-and-omega}
\end{equation}

Interestingly, since the maximum value of $\omega$ is obtained when $t_c$ is minimum at $u=0$, we have the upper bound
\beq 
\omega \leq \frac{p}{(p + \lambda \bar{X})(C_1 + \Phi^{-1}(1 - \theta_1) \sqrt{C_2})}. 
\eeq

If we neglect the constant term and the term with square root for $t_c$, then we have $\omega = \frac{1}{\Ebb[T]} \sim \frac{p \lambda \bar{X}}{u(p + \lambda \bar{X})} = \frac{p\rho}{u}$.


\section{No Energy State Information}
\label{sec:zero-bit}

In this case, since we do not have any information on the battery state, we need to rely on the statistical constraints (\ref{eqn:energy-outage}). Unlike other cases, here $T$ is a control parameter. Let the harvest duration be $t_c$, as given by (\ref{eqn:switch-on-time}), and the consumption duration be $T-t_c>0$. Here, we do not concern ourself with complete discharge of the battery. Rather, we focus on the consumption of fixed $u$ amount of energy within the consumption phase. Once this amount of energy is consumed, the system reverts to the harvest phase. Thus, some excess energy may remain in the battery after the consume phase. For simplicity, we will neglect the excess energy in the analysis. This is equivalent to assuming that any excess energy after a complete harvest-consume cycle is wasted or dissipated unproductively.

Since the consumed power is maintained at fixed $p$, the consumed energy is $u = p t_d = p \rho T$. Substituting this value of $u$ in (\ref{eqn:switch-on-time}), dividing both sides by $T$, and using the fact that $t_c/T = 1 - t_d/T = 1- \rho$,  we  have $1 - \rho = d + \sqrt{c+b \rho} + a \rho$, where $a = p/\lambda\bar{X}$, $b=p(\gamma \Phi^{-1}(1-\theta_1))^2/\bar{X}^3 T$, $c=C_2 (\Phi^{-1}(1-\theta_1)/T)^2$, and $d=C_1/T$. Here, all the constants $a,b,c,d \geq 0$. Completing the square and solving $\rho$, we obtain
\beq
\rho = \frac{2(1+a)(1-d) + b \pm \sqrt{b^2 + 4(1+a)((1+a)c + b(1-d))}}{2(1+a)^2}.
\label{eqn:duty-cycle}
\eeq

When $T \to \infty$, the parameters $b\to 0$, $c \to 0$, and $d \to 0$, thus $\rho \to \frac{1}{1+a} =  \frac{\lambda \bar{X}}{p + \lambda \bar{X}}$, regardless of the value of $\theta_1$.
  
\subsection{Feasibility Conditions}
For the solution $\rho$ to be feasible, $\rho$ should be within the interval $(0,1)$. Thus we need to check the conditions when $\rho > 0$ and $\rho < 1$. 

For $\rho > 0$, from (\ref{eqn:duty-cycle}), after some simplification, we obtain the condition $(a+1)^2 (c - (d-1)^2) > 0$. Since $a+1$ is always positive, we have the desired condition $c > (d-1)^2$.
Substituting the definitions of $c$ and $d$, we find that this condition reduces to $T > t_{c,\min}$, where $t_{c,\min}$ is as given in Section \ref{sec:one-bit}. If the terms $c$ and $d$ were neglected, then the condition would reduce to $a+1>0$, which is always true.

For $\rho < 1$, from (\ref{eqn:duty-cycle}) after some simplification, we obtain we have the desired condition $(a+d)^2 > b+c$.
Substituting the expressions for $b, c,$ and $d$, results in the condition $f(T) > 0$, where the function $f(T) = K T^2 +  LT + M$, with coefficients $K = a^2$, $L = 2 a C_1 + p \Phi^{-1}(1-\theta_1)^2/\bar{X}^3$, and $M =  C_1^2 - C_2 \Phi^{-1}(1-\theta_1)^2$.  The condition $f(T) > 0$ is satisfied for any $T$ if the discriminant of $f(T)$ is negative. That is, if $L^2 - 4KM < 0$. When this is not the case, we have $T > T_+$, where $T_+$ is the largest root of $f(T) = 0$ given by $T_+ = (-L + \sqrt{L^2 - 4KM})/2K$. Had we neglected $c$ and $d$, the condition $(a+d)^2 > b+c$ 
would have simplify to $a^2 > b$; and substituting the expression for $a$ and $b$, and solving for $T$ would have given us $T > \frac{\lambda^2 \gamma^2 }{p \bar{X}} [\Phi^{-1}(1-\theta_1)]^2 $. 

Hence, we have the lower bound on $T$ as $T > \max (t_{c,\min},T_+ )$ and the upper bound on $\omega$ as
\beq 
\omega < \frac{1}{\max( t_{c,\min},T_+)}.
\eeq

\subsection{Possible Variation} 
If proper discharge is to be ensured for fixed cycle period $T$, then allowed discharge time is $t_d = T - t_c$. Thus, we have from (\ref{eqn:discharge-time-distr})
\begin{align*} 
P(\tau_d \leq T - t_c) 
& = \Phi \left( \frac{pT - ( p + \lambda \bar{X}) t_c}{ \gamma \lambda^{3/2} \sqrt{t_c}} \right). 
\end{align*} 
Let the probability that battery is fully discharged by time $t_d$ be constrained at $P(\tau_d \leq t_d) = \theta_3$. Then, we have
\beq 
T = \left(1 + \frac{\lambda \bar{X}}{p} \right) t_c + \frac{\gamma \lambda^{3/2}}{p} \sqrt{t_c} \Phi^{-1}(\theta_3). 
\eeq

If we ignore the square root terms, for large $u$, we have the approximation $T \sim \left( \frac{1}{\lambda \bar{X}} + \frac{1}{p}\right) u$; and similarly, the duty cycle $\rho =1 - t_c/T$ is $\rho \sim  \frac{\lambda \bar{X}}{p + \lambda \bar{X}}$. Likewise, the cycle speed of the system is $\frac{1}{T} \sim \frac{1}{u}\left( \frac{\lambda \bar{X} p}{\lambda \bar{X} + p} \right) = \frac{\rho p}{u}$.

\begin{figure}[t]
\begin{center}
\subfloat[]{\includegraphics[width=0.70\columnwidth]{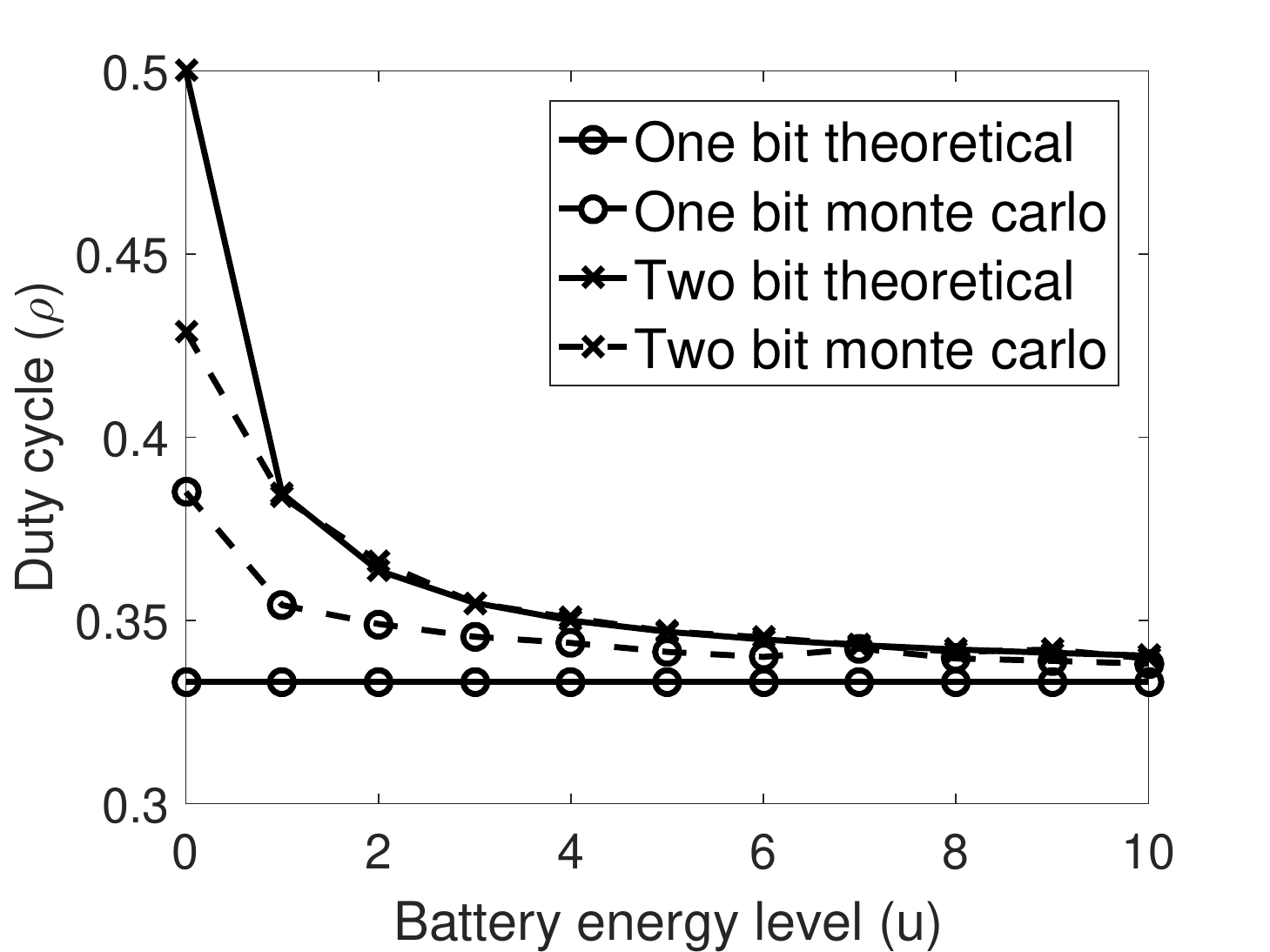}
\label{fig:dutyCycle}
}\hspace{0.001in} 
\renewcommand{\thesubfigure}{b}
\subfloat[]{\includegraphics[width=0.70\columnwidth]{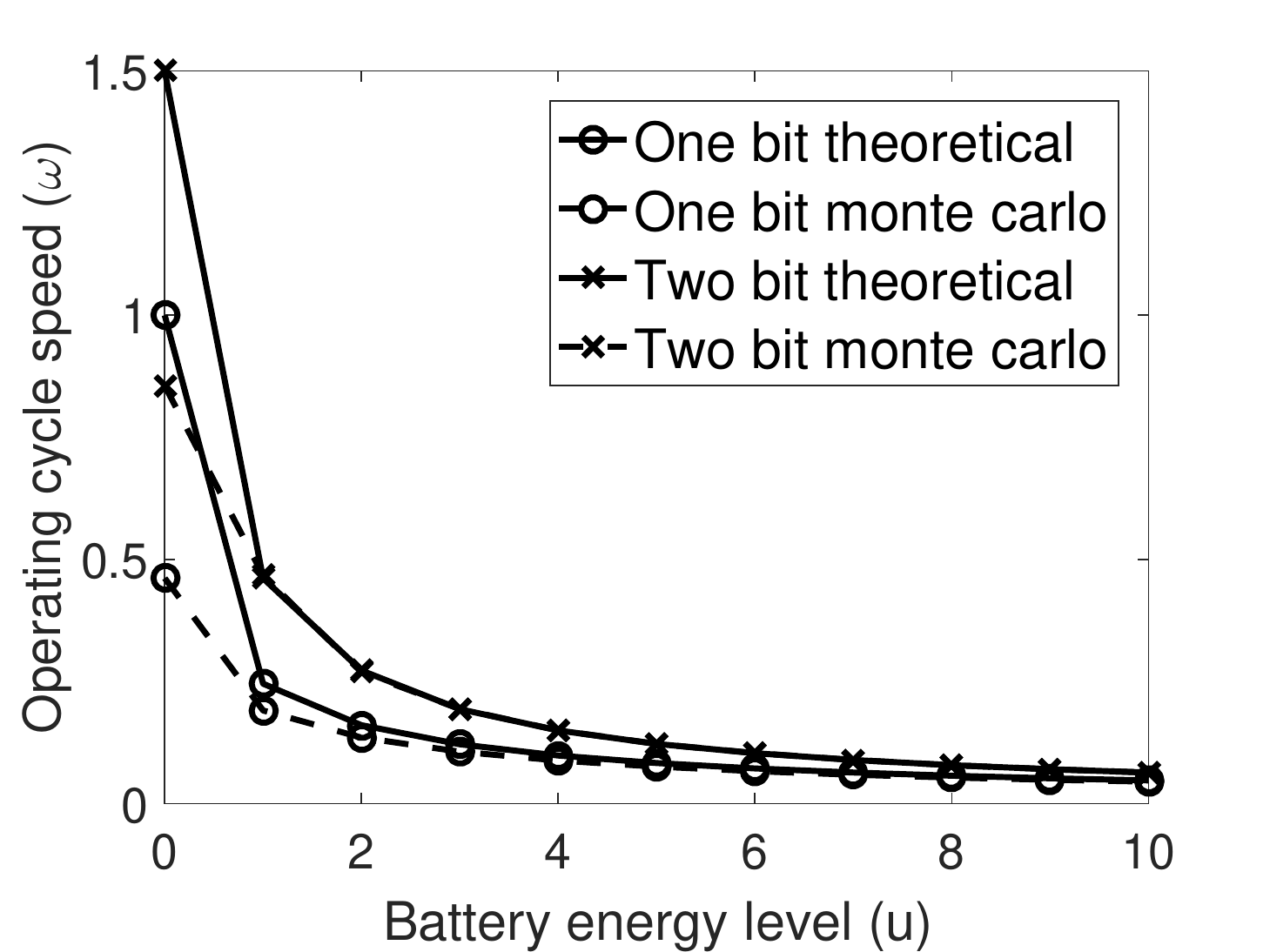}
\label{fig:frequency}%
}
\caption{\small (a) Duty cycle versus battery energy level, (b) Operating cycle speed versus battery energy level, when energy packet size $X$ and inter-arrival time $A$ are uniformly distributed.  }

\label{fig:PossionArrival}
\end{center}
\end{figure}

\section{Numerical Verification}
In this section, we verify the obtained formulas with Monte Carlo simulations for the case of two-bit and one-bit ESI. In Fig.~\ref{fig:dutyCycle} and Fig.~\ref{fig:frequency} we plot the duty cycle $\rho$ and operating cycle speed $\omega$ with respect to the threshold energy level $u$. For the two bit case, since the time $\tau_c$ and $\tau_d$ are known, it is easy to calculate the charging and discharging time, and hence the duty cycle and operating frequency. However, for one bit case, charging time $t_c$ is calculated using (\ref{eqn:switch-on-time}) and discharge time $\tau_d$ is calculated using $U(t_c)/p$. Both the energy packet size $X$ and energy arrival $A$ are assumed to follow a uniform distribution $\mathrm{U}(0,2)$, with unit mean and variance $1/3$. We assume that the power consumption $p=2$  and $\theta_1=0.1$. For a given $u$, 10,000 simulations are run to obtain a single value of $\rho$ and $\omega$. 

The theoretical expressions for $\rho$ and $\omega$ for two bit ESI are given by equations (\ref{eqn:two-bit-rho}) and (\ref{eqn:two-bit-omega}), respectively; and for one bit ESI, $\rho$ and $\omega$ are given by equations (\ref{eqn:one-bit-rho-and-omega}), respectively. From Fig.~\ref{fig:dutyCycle} and Fig.~\ref{fig:frequency}, we see that both $\rho$ and $\omega$ do not vary much for higher values of $u$. The results from the simulations match closely with the theoretical predictions. 


\section{Conclusion}
A level-triggered harvest-then-consume protocol has been proposed. The duty cycle and operating cycle speed of the system have been derived for cases when the system has two-bits, one-bit, and zero-bit of battery energy state information. Upper bounds on the system's speed have been obtained. Monte Carlo simulations have been performed to verify the obtained formulas.

\bibliographystyle{IEEE} 

\end{document}